\documentclass[aps,prc,twocolumn, superscriptaddress,floats,showpacs]{revtex4}

\usepackage{amsmath,bm}%
\usepackage{graphicx}%

\begin{document}
\title{Statistical nature of cluster emission in  nuclear liquid-vapour phase coexistence}
\author{Y. G. Ma }
\thanks{Corresponding author. Email: ygma@sinr.ac.cn}
\affiliation{Shanghai Institute of Nuclear Research, Chinese
Academy of Sciences, P.O. Box 800-204, Shanghai 201800, CHINA}
%\footnotemark \footnotetext{}
\affiliation{China Center of Advanced Science and Technology
(World Laboratory), P. O. Box 8730, Beijing 100080, CHINA}
\affiliation{Cyclotron Institute, Texas A\&M  University,
College Station, Texas 77843-3366, USA}

\author{D. D. Han}
\affiliation{Department of Electric Engineering, East China Normal University,
Shanghai 200062, CHINA }

\author{W. Q. Shen}
\affiliation{Shanghai Institute of Nuclear Research, Chinese
Academy of Sciences, P.O. Box 800-204, Shanghai 201800, CHINA}
\affiliation{China Center of Advanced Science and Technology
(World Laboratory), P. O. Box 8730, Beijing 100080, CHINA}
\affiliation{Physics Department, Ningbo University, Ningbo 315211, CHINA}

\author{ X. Z. Cai}
\affiliation{Shanghai Institute of Nuclear Research, Chinese
Academy of Sciences, P.O. Box 800-204, Shanghai 201800, CHINA}
\author{J. G. Chen}
\affiliation{Shanghai Institute of Nuclear Research, Chinese
Academy of Sciences, P.O. Box 800-204, Shanghai 201800, CHINA}
\author{ Z. J. He}
\affiliation{Shanghai Institute of Nuclear Research, Chinese
Academy of Sciences, P.O. Box 800-204, Shanghai 201800, CHINA}
\author{ J. L. Long}
\affiliation{Shanghai Institute of Nuclear Research, Chinese
Academy of Sciences, P.O. Box 800-204, Shanghai 201800, CHINA}
\author{ G. L. Ma}
\affiliation{Shanghai Institute of Nuclear Research, Chinese
Academy of Sciences, P.O. Box 800-204, Shanghai 201800, CHINA}
\author{K. Wang}
\affiliation{Shanghai Institute of Nuclear Research, Chinese
Academy of Sciences, P.O. Box 800-204, Shanghai 201800, CHINA}
\author{Y. B. Wei}
\affiliation{Shanghai Institute of Nuclear Research, Chinese
Academy of Sciences, P.O. Box 800-204, Shanghai 201800, CHINA}
\author{ L. P. Yu}
\affiliation{Shanghai Institute of Nuclear Research, Chinese
Academy of Sciences, P.O. Box 800-204, Shanghai 201800, CHINA}
\author{  H. Y. Zhang}
\affiliation{Shanghai Institute of Nuclear Research, Chinese
Academy of Sciences, P.O. Box 800-204, Shanghai 201800, CHINA}
\author{  C. Zhong}
\affiliation{Shanghai Institute of Nuclear Research, Chinese
Academy of Sciences, P.O. Box 800-204, Shanghai 201800, CHINA}
\author{X. F. Zhou}
\affiliation{Shanghai Institute of Nuclear Research, Chinese
Academy of Sciences, P.O. Box 800-204, Shanghai 201800, CHINA}
\author{ Z. Y. Zhu}
\affiliation{Shanghai Institute of Nuclear Research, Chinese
Academy of Sciences, P.O. Box 800-204, Shanghai 201800, CHINA}

%\thanks{}
%\email{}

\date{\today}
\begin{abstract}
The emission of nuclear clusters  is investigated within the
framework of isospin dependent lattice gas model and classical
molecular dynamics model. It is found that the emission of
individual cluster which is heavier than proton  is almost
Poissonian except near the transition temperature at which
the system is leaving the liquid-vapor phase coexistence
and the thermal scaling is observed by  the linear Arrhenius plots
which is made from the average multiplicity of each cluster versus
the  inverse of temperature in the liquid vapor phase coexistence.
The slopes of the Arrhenius plots, {\it i.e.} the
"emission barriers", are extracted as a function of the
mass or charge number and fitted by the formula embodied with the
contributions of the surface energy and Coulomb interaction. The
good agreements are obtained in comparison with the data for  low
energy conditional barriers. In addition, the possible influences
of the source size, Coulomb interaction and "freeze-out" density
and related  physical implications are discussed.
\end{abstract}

\maketitle

\section{Introduction}
Fragmentation phenomenon is an important subject in many different
research fields, such as in nuclear physics  and cluster physics etc.
It is found that fragmentation phenomenon has a close relation to
 liquid-to-gas phase transition in
heavy ion collision physics \cite{Gilk94,Poch95,Mayg97,Agos,ISIS}
 and cluster physics \cite{Schmidt,Gobet}.
In heavy ion collisions, the fragmentation can occur from a short-lived
hot excited nuclei with moderate temperature which can be formed by
the collisions between heavy ions at low-intermediate energy.
Usually the hot nuclei finally de-excite by the different decay modes,
such as the light particle evaporation and the emission of
multiple intermediate mass fragment ($IMF$), $i.e.$
multifragmentation. Even though extensive studies on fragmentation
have been carried out experimentally and theoretically,
it is still an open question to clarify definitely whether
the fragmentation is statistical or dynamical, sequential or
simultaneous. Among such efforts, the reducibility and thermal scaling
in multiple fragment emission process have been explored \cite{More,Tso,More97}
and seems to show one a possible interpretation picture to
look and understand the multifragmentation. Originally,
Moretto ${\it et\ al.}$ observed that the experimental $Z$-integrated fragment
multiplicity distributions $P_n^m$ are binomially distributed,
$i.e.$
\begin{equation}
P_n^m(p) = \frac{m!}{n!(m-n)!}p^n(1-p)^{m-n},
\end{equation}
in each transverse energy ($E_t$) window, where $n$ is the number of emitted
fragments and $m$ is interpreted as the number of times the system
tries to emit a fragment. The probability of emitting $n$
fragments can be reduced to a single-particle emission probability
$p$ which gives linear Arrhenius plots ($i.e.$ excitation
functions) when $ln(1/p)$ is plotted vs 1/$\sqrt{E_t}$. By
assuming a linear relationship between $\sqrt{E_t}$ and
temperature $T$, Moretto {\it et\ al.} proposed that
the linearity of the observed $ln(1/p)$ vs
1/$\sqrt{E_t}$  can be interpreted as a thermal scaling of the
multifragment process \cite{More,Tso,More97}. In this case, these
linear Arrhenius plots suggest that $p$ has the Boltzman form
$p \propto e^{-B/T}$ with  a common fragment barrier $B$.
However, since the binomial decomposition has been performed on the
$Z$-integrated multiplicities, typically associated with $3 \leq Z
\leq 20$,  the Arrhenius plot generated with the resulting one
fragment probability $p$ is an average over a range of $Z$ values.
However, some debates and comments on the above binomial distribution and
the thermal scaling were also raised \cite{Tsang,Toke,Wieloch,Botvina}.

Later, instead of analyzing for $Z$-integrated multiplicities, the
behavior of individual fragment species of a given $Z$ for higher
resolution experimental data was investigated and found that the
$n$-fragment multiplicities $P(n)$ obey a nearly Poissonian
distribution \cite{Beau,More99},
\begin{equation}
P(n) = \frac{\langle n\rangle^n e^{-\langle n\rangle}}{n!},
\end{equation}
where $n$ is the number of fragments of a given Z and the average
value $\langle n\rangle$  is a function of the total transverse energy $E_t$,
and were thus reducible to a single-fragment probability
proportional to the average value $\langle n\rangle$ for each $Z$. Similarly
the $\langle n\rangle$ is found to be proportional to $e^{-B/T}$ providing
that $T \propto \sqrt{E_t}$, $i.e.$ there exists also a thermal
scaling law.  More recently, the common features of Poissonian
reducibility and thermal scaling can also be revealed in
percolation and the Fisher droplet model \cite{Elliott,Elliott02}.

In the present work, we would like to make a theoretical
re-examination on the Poissonian reducibility and its thermal
scaling rather than the bionomial reducibility and its thermal
scaling. Unlike experiment, we will adopt the true temperature as
model input to check the Poissonian reducibility and thermal scaling
in the frameworks of the isospin dependent lattice gas model
(I-LGM) \cite{Jpan95,Jpan96} and followed by the isospin dependent
classical molecular dynamics (I-CMD)  of Das Gupta and Pan
\cite{Jpan98}. By investigating the variances and
average multiplicities of cluster multiplicity distributions  as a
function of temperature, we will illustrate that the Poissonian
reducibility and its thermal scaling is valid for the fragment
emission in the low temperature side in the framework of the above
thermal equilibrium models.

The paper is organized as follows. Firstly, we introduce the
models of I-LGM and I-CMD in Sec. II; In Sec.III, the results and
discussions are presented.  First we show some results to support
that there exists a Poissonian reducibility in the cluster production
away from the liquid gas phase transition by investigating the
ratio of the dispersion of multiplicity distribution to its mean
value and the Poissonian fit to the multiplicity distribution of
individual clusters. Second we plot the Arrhenius-type plots and
find the thermal scaling is valid only in the liquid vapor coexistence
phase. Further we extract the "emission barrier"
sorted by the different mass number, light
isotope, and charge number in I-LGM and I-CMD and use the formula
embodied with the contributions of the surface energy and Coulomb
interaction to make a systematic fit. The dependence of the model,
the source size, Coulomb interaction and "freeze-out"  density are
presented. Finally, a summary and outlook is given in Sec. IV.

\section{Description of Models}

Originally, the lattice gas model was developed to describe the
liquid-gas phase transition for atomic system by Lee and Yang
\cite{Yang52}. The same model has already been applied to nuclear
physics for isospin symmetrical systems in the grand canonical
ensemble \cite{Biro86} with a sampling of the canonical ensemble
\cite{Jpan95,Jpan96,Jpan98,Mull97,Camp97,Gulm98,Carmona98,Ma99,Ma99PRC,Ma99CPL},
and also for isospin asymmetrical nuclear matter in the mean field
approximation \cite{Sray97}. In this work, we will adopt the
lattice gas model which was developed by Das Gupta ${\it et\
al.}$ \cite{Jpan95,Jpan96}. In addition, a classical molecular
dynamical model of Das Gupta ${\it et\ al.}$ \cite{Jpan98}
is also used to compare with the results of lattice gas model. For
completeness of the paper, here we make a brief description for
the models.

In the lattice gas  model, $A$ (= $N + Z$) nucleons with an occupation number
$s_i$ which is defined $s_i$ = 1 (-1) for a proton (neutron) or $s_i$ = 0 for
a vacancy, are placed on the $L$ sites of lattice. Nucleons in the nearest
neighboring sites have interaction with an energy $\epsilon_{s_i s_j}$.
The Hamiltonian is written as
\begin{equation}
E = \sum_{i=1}^{A} \frac{P_i^2}{2m} - \sum_{i < j} \epsilon_{s_i s_j}s_i s_j ,
\end{equation}
where $P_i$ is the momentum of the nucleon and $m$ is its mass.
The interaction constant $\epsilon_{s_i s_j}$ is chosen to be isospin dependent
and be fixed to reproduce the binding energy of the nuclei \cite{Jpan98}:
\begin{eqnarray}
 \epsilon_{nn} \ &=&\ \epsilon_{pp} \ = \ 0. MeV \nonumber , \\
 \epsilon_{pn} \ &=&\ - 5.33 MeV.
\end{eqnarray}
 Three-dimension cubic lattice with $L$ sites is used which
results in $\rho_f$ = $\frac{A}{L} \rho_0$ of an assumed "freeze-out" density
of disassembling system, in which $\rho_0$ is the normal nuclear density.
The disassembly of the system is to be calculated at $\rho_f$, beyond
which nucleons are too far apart to interact.  Nucleons are put into
lattice by Monte Carlo Metropolis sampling. Once the nucleons
have been placed we also ascribe to each of them a momentum by Monte Carlo
samplings of Maxwell-Boltzmann distribution.

Once this is done the I-LGM immediately gives the cluster distribution
using the rule that two nucleons are part of the same cluster if
\begin{equation}
 P_r^2/2\mu - \epsilon_{s_i s_j}s_i s_j < 0 ,
\end{equation}
where $P_r$ is the relative momentum of two nucleons and $\mu$ is their
reduced mass. This prescription is evidenced to be similar to the Coniglio-Klein's
prescription \cite{Coni80} in condensed matter physics and be valid in I-LGM
\cite{Camp97,Jpan96,Jpan95,Gulm98}. To calculate clusters using I-CMD we
propagate the particles from the initial configuration for a long time under
the influence of the chosen force. The form of the force is chosen to be also
isospin dependent in order to compare with the results of I-LGM. The potential
for unlike nucleons is \cite{Jpan96,Jpan98,Stillinger}
\begin{eqnarray}
 v_{\rm n p}(r) (\frac{r}{r_0}<a)\ &=&\ A\left[B(\frac{r_0}{r})^p-(\frac{r_0}{r})^q\right]\nonumber
    exp({\frac{1}{\frac{r}{r_0}-a}}), \\
v_{\rm  n p}(r) (\frac{r}{r_0}>a)\ &=&\ 0.
\label{pot}
\end{eqnarray}
In the above, $r_0 = 1.842 fm$ is the distance between the centers of two adjacent
 cubes. The parameters of the potentials are $p$ = 2, $q$ = 1, $a$ = 1.3,
$B$ = 0.924, and $A$ = 1966 MeV. With these parameters the
potential is minimum at $r_0$ with the value -5.33 MeV, is zero
when the nucleons are more than 1.3$r_0$ apart and becomes
stronger repulsive when $r$ is significantly less than $r_0$. The
potential for like nucleons is written as
\begin{eqnarray}
v_{\rm p p}(r) ( r < r_0 )\ &=&\  v_{\rm n p}(r)- v_{\rm n p}(r_0)\nonumber , \\
v_{\rm p p}(r) ( r > r_0 )\ &=&\ 0.
\end{eqnarray}
This means there is a repulsive core which goes to zero at $r_0$
and is zero afterwards. It is consistent with the fact that we do
not put two like nucleons in the same cube. Essentially the
classical molecular dynamics  is based on the idea that the
equation of state of classical particles interacting through
attractive and repulsive Yukawa or Lennard-Jones potentials leads
to an equation of state similar to a Van de Waals one. This kind
of approach was first proposed in Ref. \cite{Lenk} to treat the
nuclear collision with the similar potential as used in this
paper.  Extensive studies using the similar CMD to explore the
liquid gas phase transition and fluctuations were also performed
in some works, such as Ref. \cite{Pratt,Latora,Dorso,Bonasera}.

The system evolves for a long time from the initial configuration obtained
by the lattice gas model under the influence of the above potential.
At asymptotic times the clusters are
easily recognized. The cluster distribution and the quantities based on it in
the two models can now be compared. In the case of proton-proton interactions,
the Coulomb interaction can also be added separately and compared with the
cases where  the Coulomb effects are ignored.

\section{Results and Discussions}

In this paper we  choose the medium size nuclei $^{129}$Xe  as a
main example to analyze the behavior of individual fragment
emission during nuclear disassembly with the helps of I-LGM and
I-CMD. In addition, the systems with $A_{sys}$ = 80 ($Z_{sys}$ =
33) and 274 ($Z_{sys}$ = 114) are also studied to investigate the
possible source size dependence. One part of this work has been
reported in Ref.\cite{Ma01CPL} for $^{129}$Xe in the I-LGM
calculation with a fixed "freeze-out" density. In this work,
a lot of new calculations are largely included.

In both model calculations, the "freeze-out" density $\rho_f$ is
mostly chosen to be about 0.38 $\rho_0$,
since the experimental data can be best
fitted by $\rho_f$ between 0.3$\rho_0$ and 0.4$\rho_0$ in the
previous LGM calculations \cite{Jpan95,Beau96}, which corresponds
to  $7^3$ cubic lattice is used for Xe, $6^3$ for $A_{sys}$ = 80
and $9^3$ for $A_{sys}$ = 274. Under the condition of the fixed
"freeze-out" density, the only input parameter of the models is the
temperature $T$. In the I-LGM case, $\rho_f$ can be thought as the
"freeze-out" density but in the I-CMD case $\rho_f$ is, strictly
speaking, not a "freeze-out" density but merely defines the starting
point for time evolution. However since classical evolution of a
many particle system is entirely deterministic, the initialization
does have in it all the information of the asymptotic cluster
distribution, we will continue to call $\rho_f$ as the "freeze-out"
density. 1000 events are simulated for each $T$ which ensures
enough statistics.

\subsection{Poissonian Reducibility}

\begin{figure}
%\vspace{-0.1truein}
\includegraphics[scale=0.40]{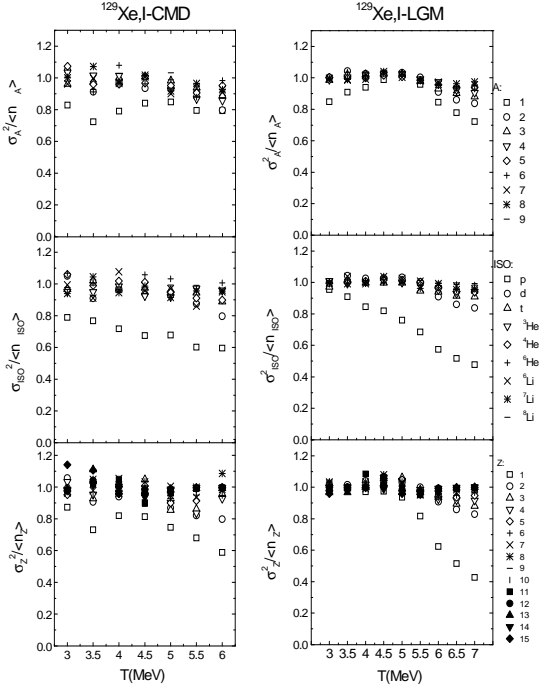}
%\vspace{-0.5truein}
\caption{\footnotesize The ratio of ${\sigma_i}^2/\langle n_i\rangle$ for the
clusters classified with mass, light isotope mass and atomic
number as a function of temperature. The left panels are for the
I-CMD calculation and the right for I-LGM with Coulomb. The
symbols are illustrated on the figure.}
\label{sgm_n_ratio}
\end{figure}

One of the basic characters of the Poisson distribution Eq.(2) is
the ratio $\sigma_{n_i}^2/\langle n_i\rangle \rightarrow 1$ where
$\sigma_{n_i}^2$ is the variance  of the multiplicity distribution
and $\langle n_i\rangle$ is the mean value of the multiplicity distribution.
The first step we are showing is this ratio. We obtain these
ratios for clusters classified with different masses ($A$), light
isotopes ($ISO$) and atomic numbers ($Z$) for the disassembly of
$^{129}Xe$ as a function of temperature in the framework of I-LGM
and I-CMD with Coulomb in Fig.~\ref{sgm_n_ratio}. Obviously, most
of the ratios are close to one, which indicates that basically
these cluster production obeys the Poisson distributions, $i.e.$ a
cluster is formed independently from one another. Of course, we
also notice that the values of protons are almost lower than the
unique, $i.e.$ it is narrower than the Poisson distribution. This
could be due to protons can be easily separated from some unstable
multi-nucleon clusters. In this case, this kind of proton
production is obviously related to the parent cluster and then a
narrower distribution of mixed protons could reveal. On the
contrary, some points are slightly larger than 1 and this behavior
becomes obvious in the mediate temperature range, which is happening
because at this temperature the system is leaving the liquid vapor
phase coexistence  \cite{Jpan95},
 where the large fluctuation \cite{Ma01} makes the Poissonian
reducibility broke-down.

\begin{figure}
\includegraphics[scale=0.40]{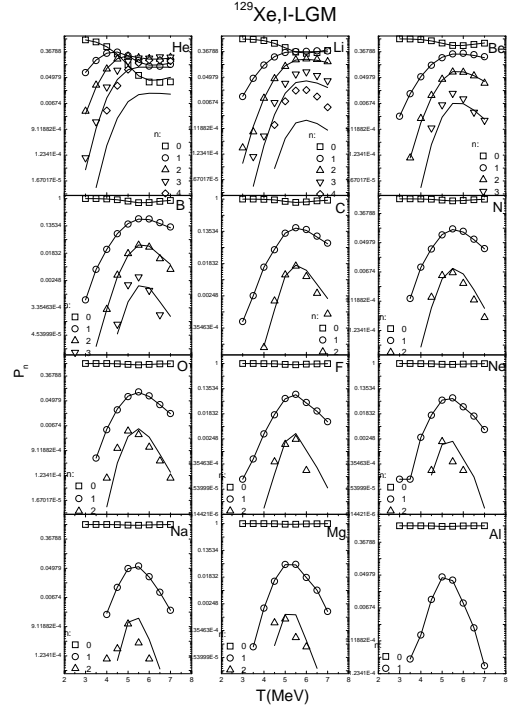}
\caption{\footnotesize The excitation functions of $n$-multiple
cluster emission probability ($P_n$) for elements
with $Z \geq 2$ emission from the source $^{129}Xe$ in the I-LGM
calculation. The lines show the expected values in different
temperature  with the Poisson assumption according to  Eq.(2).}
\label{Pn}
\end{figure}

Besides the above Poisson condition is basically sustained,
 the excitation function of the average multiplicity of
$n$-multiple  individual cluster emission can be well fitted
with the Poisson distribution.  For some examples,
Fig.~\ref{Pn}  shows the quality of the Poissonian
fits to the  average multiplicity of
$n$-multiple  individual cluster emission in the different temperature
for $^{129}Xe$ in the I-LGM case. In each panel of this figure, we
first plot the probability of $n$-multiple emission species ( P($n$) ) as a
function of temperature (as shown by the open symbols),
and then we connect the Poisson probabilities in the different temperature
as lines  in terms of  Eq.(2) due to we know $n$ and its average value $\langle n\rangle$
over all $n$-multiple emission in each temperature (as shown by the
lines).  Obviously, these Poissonian fits are quite good for
almost species with $Z \geq 2$ over the entire range of $T$.
The similar  good Poissonian fit is overall obtained in the cases of I-CMD.
As a consequence, the Poissonian reducibility is valid
in the thermal-equilibrium lattice gas model or molecular dynamics,
which illustrates that the cluster production is almost
independent each other in the studied models.

\subsection{Thermal Scaling}

\begin{figure}
\includegraphics[scale=0.40]{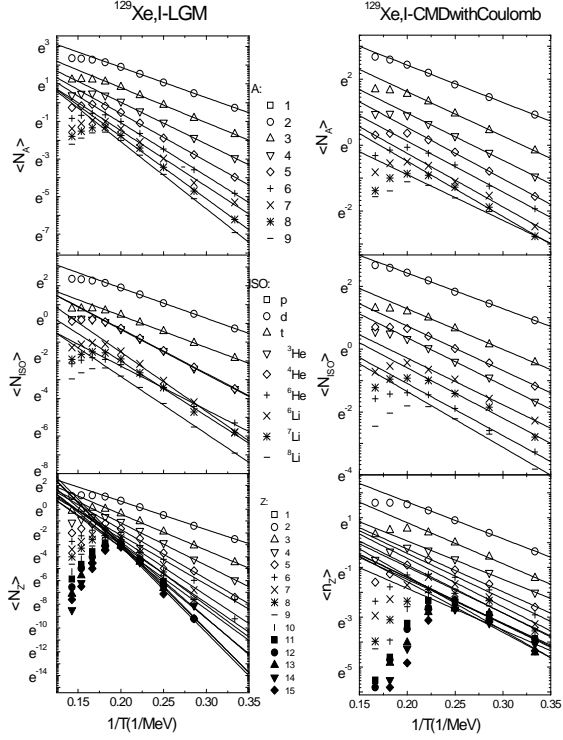}
\caption{\footnotesize Arrenhius-type plot: the average yield per
event of different clusters classified with $A$ (top), $ISO$
(middle) and $Z$ (bottom) as a function of $1/T$. The left panels
show the results with I-LGM calculation and the right present the
results with  I-CMD including Coulomb force. The solid lines are
fits to the calculations using a Boltzman factor for $\langle n_i\rangle$.
 The symbols are illustrated on the figure.}
\label{thermal}
\end{figure}

Naturally, we want to know if there exists a  thermal scaling law
in the thermal-equilibrium LGM and CMD models. To this end,
the temperature dependence of the mean yield ($\langle n\rangle$) of individual
clusters is investigated. In order to compare a recent well-known
thermal Arrhenius-type plot in nuclear multifragmentation phenomenon
\cite{More97,Tsang}, we plot $ln \langle n\rangle$ versus $1/T$.
Fig.~\ref{thermal} gives a family of these plots for the disassembly
of $^{129}Xe$ within the framework of I-LGM (left panels) and I-CMD
with Coulomb interaction (right panels). Again, as Fig.~\ref{sgm_n_ratio},
the clusters are classified according to their masses (upper panels),
the light isotopes (the middle panels) and the charge numbers (the
lower  panels).  For all the panels, the obtained Arrhenius
plots are  linear for the lower $T$ side, and their slopes
generally increase with increasing $A$ or $Z$ value.
At these temperatures, one can anticipate one large
fragment surrounded by many small clusters. However, contrary
tendency reveals in the high $T$ side where $ln \langle n\rangle$ increases
with $1/T$, $i.e.$ decreases with increasing $T$. In this case,
nuclear Arrhenius plots of $\langle n\rangle$ with $1/T$ are
invalid but the Poissonian reducibility still remains (see Fig.~\ref{Pn}).
This behavior of $\langle n\rangle$ at higher $T$ is related to the branch of
the fall of the multiplicity of $IMF$ ($N_{IMF}$) with $T$ where
the disassembling system is in vaporization \cite{Ogil,Tsang93,Ma95}
and hence only the lightest clusters are dominated and the
liquid residue is vanishing with increasing $T$.
Above the temperature at which the Arrenhius-type plot begins to
deviate from the linearity,  the fragments
are no longer in coexistence with the big liquid drop. In this case,
the supply of droplets to the vapor has been exhausted and the system is only
in a single (gas) phase.
In other words, the Arrenhius law looks valid only
when the disassembling system is in  coexistence phase of liquid and vapor.
Recognizing this phenomenon, in the following
sections we only focus on the branch of lower temperature ($i.e.$ in
phase coexistence) where the thermal scaling exists to discuss the
Arrhenius law and their slopes.

\subsection{"Emission Barriers"}

\subsubsection{Model dependence}

\begin{figure}
\includegraphics[scale=0.40]{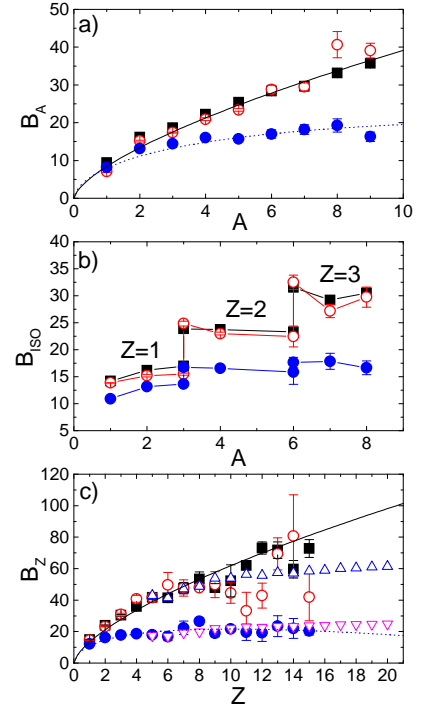}
\caption{\footnotesize The "emission barrier" extracted from the
Arrhenius-type plots as a function of cluster mass (top), isotopic
mass (middle) or cluster charge (bottom) in the cases of I-LGM
(solid squares), I-CMD without Coulomb (open circles) and with
Coulomb (solid circles). The unit of the "emission barrier" is MeV
throughout this paper. The solid lines are fits with the Eq. (8)
or (9), and the dot-dashed lines represent the fits with the Eq.
(10) or (11). The open up-triangles and down-triangles are low
energy conditional barriers for $^{82}Kr + ^{12}C \rightarrow
^{94}Mo$ from Ref.\cite{Jing} normalized to $Z$ = 6 for I-LGM case
and I-CMD case, respectively.}
\label{barrier_model}
\end{figure}

From Fig.~\ref{thermal} the slope parameter can be directly  extracted in
the lower $T$ side as a function of $Z$ or $A$. In Ref.
\cite{Beau} Moretto ${\it et\ al.}$ has interpreted
these slope parameters as "emission barriers" of specific
individual fragments. This "emission barrier" is usually
associated with pictures of sequent emission from a
compound-like nucleus. Fig.~\ref{barrier_model}
gives the "emission barrier" of individual fragments with different
$A$, $ISO$ and $Z$ in the framework of I-LGM, I-CMD with/without
Coulomb interaction. The error bar in the figure represents the
error in the extraction of the slope parameter.
 The first indication from this figure is that the "emission
barrier" in the I-LGM case is the nearly same as the  I-CMD case
without Coulomb force, which supports  that I-LGM is equivalent to
I-CMD without Coulomb interaction rather well when the nuclear
potential parameter is suitably chosen, but I-LGM is a quick
model to analyze the behavior of nuclear dissociations. The
inclusion of long-range Coulomb interaction makes the "emission
barrier" of individual fragments much lower since the repulsion of
Coulomb force reduces the attractive role of potential and hence
make clusters escape easily. The second indication is that the
"emission barrier" increases with $A$ ($Z$) at low $A$ ($Z$)
values and tend to be saturated at high $A$ ($Z$) ones. Similar
experimental trends have been observed  for individual fragments
with different $Z$ in Ref.\cite{Beau,Jing} or different $A$ in
Ref.\cite{Elliott}. To make a comparison with the experimental
data, we plot the data for low energy conditional barriers for
$^{82}Kr + ^{12}C \rightarrow ^{94}Mo$ \cite{Jing} in the same
Fig.~\ref{barrier_model}c. Since the unknown quantitative relation
of the temperature between the present model and the data, we
normalize the above conditional barriers to $Z$ = 6 for I-LGM case
and I-CMD case, respectively, as done in Ref.\cite{Beau}. Overall
good agreements between the calculation and data are obtained. We
noticed that the middle panel of Fig.~\ref{barrier_model} shows
that insensitivity of "emission barrier" of $ISO$ on $A$ in the
fixed atomic number $Z$, which indicates that the $Z$ dependence
of barrier is perhaps more intrinsic and the $A$ dependence is
basically due to the average effect over the species with the same
$A$ but different $Z$.

\subsubsection{Source size and Coulomb interaction dependence}

On the origin of these barrier, the surface energy and Coulomb
energy would play the roles. If the cluster emission is mainly
controlled by its surface energy, it would suggest barriers
proportional to $Z^{2/3} (A^{2/3})$.
In the case of I-LGM and I-CMD without Coulomb, we can try to
fit the barrier for the particles with different mass number by

\begin{equation}
B_{Coul. off} = c_1 \times {A_i}^{2/3},
\end{equation}

or for the particles with different charge number by

\begin{equation}
B_{Coul. off} = c_1 \times ( (A/Z)_{fit} * Z_i)^{2/3},
\end{equation}
where $(A/Z)_{fit}$ is a fit coefficient of A/Z for emitted
particles, and $A_i$ ($Z_i$) is the mass (charge) of particle.
$c_1$ is the fit constant for surface energy term. The solid line
in the Fig.~\ref{barrier_model}a is a function of Eq.(8) with
$c_1$ = 8.469 and the solid line in the Fig.~\ref{barrier_model}c
is a function of Eq.(9) with $c_1$ = 8.469 and $(A/Z)_{fit}$ =
1.866. These excellent fits imply that the surface energy play a
major role in controlling the cluster emission when the long range
Coulomb force is not considered. However for the cluster emission
with the Coulomb field, we can assumed that the barrier is mainly
constituted by the surface energy term and an additional Coulomb
term as

\begin{widetext}
\begin{equation}
B_{Coul. on} = c_2 \times  A_i^{2/3} -
              \frac{1.44 \times A_i/(A/Z)_{fit} \times Z_{res}}
{r_{Coul} ({A_i}^{1/3} + ((A/Z)_{fit}*Z_{res})^{1/3})}
\end{equation}
\end{widetext}
for the particles classified with different mass number, or
\begin{widetext}
\begin{equation}
B_{Coul. on}  =  c_2 \times  ( (A/Z)_{fit} * Z_i )^{2/3}
  - \frac{1.44 \times Z_i \times Z_{res}}
{r_{Coul} ( (Z_i*(A/Z)_{fit})^{1/3} + (Z_{res}*(A/Z)_{fit})^{1/3})}
\end{equation}
\end{widetext}
for the particles  classified with different charge  number, where
$c_2$ is a fit constant for surface term and  $r_{Coul}$ is chosen
to be 1.22 fm. $Z_{res}$  is a fitted average charge number of the
residue. $(A/Z)_{fit}$ is chosen to be 1.866, as taken from the
fits for I-LGM. The overall fits for  $A$ and $Z$ dependent
barrier in the case of I-CMD with Coulomb force give $c_2$ =
12.921 and $Z_{res}$ $\sim$ 41 with the dot-dashed line in
Fig.~\ref{barrier_model}a and 4c. The excellent fit supports that
the Coulomb energy plays another important role in the
cluster emission.

\begin{figure}
\includegraphics[scale=0.40]{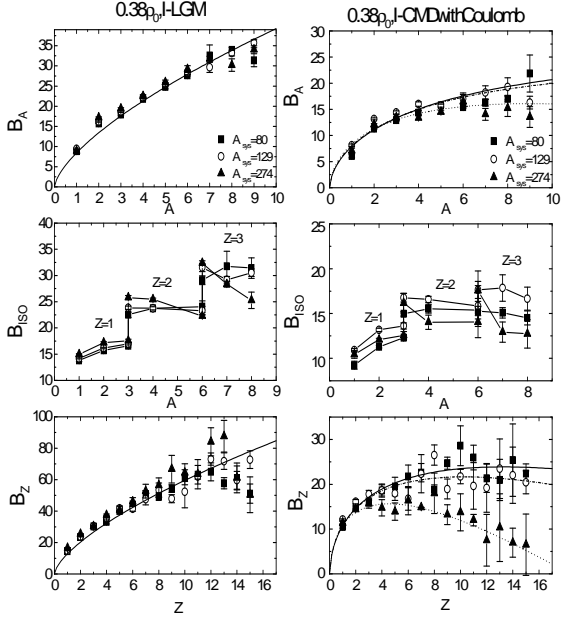}
\caption{\footnotesize The source size dependence of the  emission
barriers for the different clusters classified with mass (top),
isotopic mass (middle) or cluster charge (bottom) from in the
cases of I-LGM (left panels), I-CMD with Coulomb (right panels). The
lines in the left panel are fits with the Eq. (8) or (9), and the
solid, dot-dashed and dotted line in the right panel represents
the fits to the "emission barrier" of $A_{sys}$ = 80, 129 and 274,
respectively, with the Eq. (10) or (11).}
\label{barrier_A}
\end{figure}

In the case of I-LGM and I-CMD without Coulomb, one would expect
the barrier for each $Z$ ($A$) to be nearly independent of the
system studied if only the surface energy is substantial to the
"emission barrier". The left panels of the Fig.~\ref{barrier_A}  show the
results for $B_A$, $B_{ISO}$ and $B_Z$ for three different systems
in the I-LGM case. The same "freeze-out" density of 0.38$\rho_0$ and
the same $N/Z$ is chosen for the systems of $A_{sys}$ = 80 and
$A_{sys}$ = 274. Actually, it appears to have no obvious
dependence of "emission barrier" on source size as expected for the
role of surface energy. The solid line in the figure is the same
as in Fig.~\ref{barrier_model}.   However, when the long-range
Coulomb interaction is considered, the "emission barrier" reveals
a source size dependence.
The right panels of Fig.~\ref{barrier_A} give the "emission barriers" $B_A$,
$B_{ISO}$ and $B_Z$ in the case of I-CMD with Coulomb force. It
looks that the barrier increase with the decreasing of charge of
system, which can be explained with the Eq. (10) and (11) where
the decreasing of the residue $Z_{res}$ will result in the
decreasing of the Coulomb barrier and hence the increasing of the
"emission barrier". The lines represent the fits with the Eq. (10)
and (11) for three different mass systems.

\subsubsection{"Freeze-out" density dependence}

\begin{figure}
\includegraphics[scale=0.40]{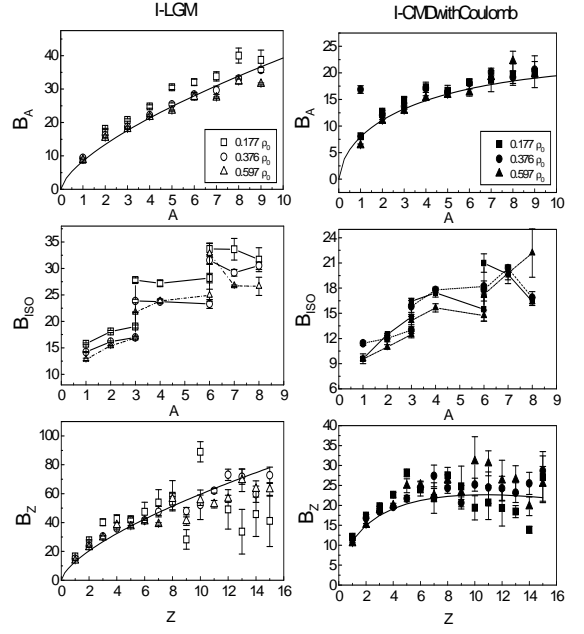}
\caption{\footnotesize The  "emission barrier" of $^{129}$Xe for the
different clusters classified with mass (top), isotopic mass
(middle) or cluster charge (bottom) at the different "freeze-out"
density in the cases of I-LGM (left panels), I-CMD with Coulomb
(right panels). The lines are fits with the Eq. (8) or (9) in the
left panels and  with the Eq. (10) or (11 )  in the right panels.}
\label{barrier_rho}
\end{figure}

In the above calculations, the "freeze-out" density of systems is
fixed at $\sim$ 0.38$\rho_0$.
 Considering the "freeze-out" density is an important debating
variable in the latter stage of heavy ion   collisions,
here we will discuss the possible influence of "freeze-out"
density on the "emission barrier" of clusters. The calculations
at the "freeze-out" density  of 0.177$\rho_0$ and 0.597$\rho_0$
for $^{129}Xe$, corresponding to $9^3$ and $6^3$ cubic lattices
respectively, are supplemented  to compare.
Fig.~\ref{barrier_rho} gives the results of $B_A$, $B_{ISO}$ and $B_Z$ at
different density. It looks that there are no obvious "freeze-out"
density dependence in the both cases of I-LGM and I-CMD. This is
also consistent with that assumption  that the surface energy is
the dominant role in controlling the cluster emission.

\section{Summary and Outlook}

In conclusion, the Poissonian reducibility and thermal scaling of the
emitted clusters is explored in the lattice gas model and
molecular dynamical model of Das Gupta and Pan. It indicates of a statistical nature
of such cluster emission. The Poissonian reducibility illustrates
that the clusters are produced independently each other and stochastic.
But near
% the critical point of liquid gas phase transition,
the transition temperature at which the system
leaves the liquid vapor phase coexistence, the large fluctuation
breaks down the Poissonian reducibility. The thermal scaling exists
while the disassembling system is at coexistence phase.
In this case, one can anticipate one large
fragment (liquid) surrounded by many small clusters (vapor).
The calculations are qualitatively
consistent with the recent experimental observation of Poissonian reducibility
and thermal scaling by Moretto/Wozniak's group even though the
 system studied is different and the temperature was supposed to be
 proportional to the total transverse energy in their experiments.
 This  also supports that the
 lattice gas model and classical molecular dynamics is a useful tool
 to simulate the nuclear disassembly.

Further, based on the Arrhenius law in the liquid phase,
we extracted  the "emission barrier" for clusters with the different
mass, light isotope mass and charge number.  The good agreement with
the experimental data of the low energy conditional barrier is
obtained if the data was normalization to $Z$ = 6 for the calculation
is used. Also, the systematic fits
with the formula embodied with the surface energy and Coulomb interaction
were performed and the overall good fits were reached.
The results suggest that the cluster emission is mainly
controlled by both the surface energy
and the Coulomb interaction. In the framework of the lattice gas model and
molecular dynamics model without the Coulomb interaction, the
"emission barrier" relies on the cluster charge with the $Z^{2/3}$
($A^{2/3}$) law and it does not depend on the source size and
"freeze-out" density, which indicates that the surface energy play a
dominant role to control the cluster emission. However,
in the framework of molecular dynamics model with the Coulomb
force, the "emission barrier" will decrease strongly according to
the Eq. (10) and (11) and it decreases with the increasing of the
source size, illustrating that the Coulomb interaction also play
another weighty role to control the cluster emission.

Due to the present models are basically statistical, even for
the present molecular dynamics model since it uses I-LGM
predictions as the initial conditions for I-CMD. In this case the
I-CMD results can not be considered as independent ones. Hence,
the dynamical transport models, such as quantum molecular dynamics
\cite{QMD}, Fermion  molecular dynamics \cite{FMD} and
anti-symmetrizied molecular dynamics \cite{AMD} etc., will be very
valuable to explore the statistical and/or dynamical features of
cluster emission since these dynamical models include some
essential ingredients of nuclei and their collisions, such as
Fermi motion,  Pauli blocking, mean field effect and collision
term etc. They can produce clusters by stochastic motion without
the initial thermal equilibrium assumption and "clustering". In
fact, some studies have been carried on for investigating the
binomial scaling of cluster emission in terms of an elementary
binary decay mechanism using a simultaneous statistical
multifragmentation model and a molecular dynamics  model, such as
Ref. \cite{Donangelo}. A very different feature has been found in
that study. In the same sprit, it is also meaningful to use a
molecular dynamics model followed by a statistical decay model
\cite{Ma02} to explore the Poissionian reducibility and thermal
scaling. The work along this line is also in progress.

Finally we would like to emphasize that the present work is
based on the canonical model calculation, {\it i.e.}, the temperature
is fixed in the Monte Calro simulations, which is appropriate
for a system in contact with a large heat bath.
However, a microcanonical treatment to the multiplicity fluctuation
should be also very interesting since the energy conservation
will probably play a different role in fragment fluctuation.
Recently, Pratt and Das Gupta employed a fragment gas  model
to discuss the importance of energy
conservation (canonical vs microcanonical) in affecting these
fluctuation \cite{Pratt2}.  By investigating a correlation
coefficient which was defined as the variance of multiplicity
distribution relative to its mean in such a way that it
is positive or negative for super- or sub-Poissonian distributions,
they found that the correlation coefficient developed
a singularity at the critical point when plotted against
the temperature in canonical calculations; whereas in microcanonical
calculations the correlation coefficient reached a gentle maximum
for excitation energies in the fragmentation region.
The energy conservation ({\it i.e.} microcanonical case)
results in decreasing of multiplicity fluctuation and
the multiplicity distribution always returns to be sub-Possionian.
However, even though the singular behavior of the correlation
coefficient was muted in a microcanonical treatment, the
behavior of the correlation coefficient as a function
of the excitation energy was unique to the
process of statistical fragmentation. Of course,
this fluctuation phenomenon in different statistical
ensembles  could be model dependent. For instance,
in a recent paper \cite{Das}, Das {\it et.\ al.} showed that
 microcanonical calculation of LGM has no serious
departures from canonical results, eg., the
caloric curve and fluctuation of IMF distribution are
very similar between canonical and microcanonical treatment.
Detailed analysis and comparisons of the reducibility
and thermal scaling with the microcanonical
LGM are certainly welcome.

\acknowledgments

Authors are grateful to Prof. S. Das Gupta
 and Dr. J.C. Pan for  providing the orginal LGM and CMD codes kindly.
  This work was supported
in part by the Major State Basic Research Development Program of
China under Contract No. G200077400 and the National Science
Foundation of China under  Grant No. 10135030 and 19725521.

{}
\end{document}